# Issues related to the usage of nitrogen as carrier gas for the MOVPE growth of GaSb/InAs heterostructures on InAs pseudosubstrates

Hanno Kroencke[1,2], Agnieszka Gocalinska[1,*], Valeria Dimastrodonato[1], Marina Manganaro[1,3], Emanuele Pelucchi[1]

[1]Tyndall National Institute, Dyke Parade, Lee Maltings, Cork, Ireland

[2]current address: Helmholtz-Zentrum Berlin für Materialien und Energie GmbH Institute of Nano-architectures for Energy Conversion Hahn-Meitner-Platz 1, 14109 Berlin, Germany

[3]current address: Inst. de Astrofísica de Canarias, E-38200 La Laguna, Tenerife, Spain; Universidad de La Laguna, Dpto. Astrofísica, E-38206 La Laguna, Tenerife, Spain

**Corresponding Author**

* e-mail agnieszka.gocalinska@tyndall.ie, Phone: +353 21 420 6625

**ABSTRACT** GaSb/InAs/GaSb layer stacks have been grown on InAs metamorphic substrates (pseudosubstrates) by MOVPE, using nitrogen as major carrier gas. We demonstrate that GaSb growth by "nitrogen MOVPE" on InAs metamorphic substrates (and on InAs wafers) is possible in a very narrow range of growth parameters. As demonstrators, GaSb/InAs/GaSb structures were grown for electron mobility tests, obtaining (unintentional) 2D electron gas densities in the order of $9/5*10^{12}$ cm$^{-2}$ and mobilities up to $1.2/1.8*10^{4}$ cm$^{2}$/Vs at room and liquid nitrogen temperature respectively. We show that it is beneficial to have some hydrogen in the carrier gas mixture for GaSb growth to achieve good crystal quality, morphology and electrical properties. Furthermore, an unexpected and previously unreported decomposition process of GaSb is observed at relatively low growth temperatures under the supply of the precursors for InAs epitaxial overgrowth. This nevertheless gets suppressed at even lower growth temperatures.

1.   **Introduction**

A number of applications of antimony-based structures have been envisioned and reported for detectors and emitters in the IR region, for magnetic sensors and also for high electron mobility transistors, between the many. For these applications a nearly lattice matched substrate would be, in general, a good choice but this is not always economical or, practically, a best choice. GaSb as well as InAs substrates are not commercially available in their "insulating" version and have high costs, which hampers their industrial exploitation. As an alternative, GaAs can (and has been) used as a substrate [1, 2, 3, 4], with the extra inconvenience of strain and relaxation management and associated issues for the heavily lattice-mismatched InAs/GaSb layers. Moreover, there are significant differences in the quality of the outcomes between the epitaxial techniques reported and exploited. Most of the high quality results have been achieved by molecular beam epitaxy (MBE), for example by authors in [5] who achieved mobilities of $\mu = 2.7*10^4$ cm$^2$/Vs in an InAs/AlSb system, or by authors in [6], who measured mobilities up to $\mu_{LT} = 200*10^4$ cm$^2$/Vs in a MBE grown InAs/GaSb/AlSb system at cryogenic temperatures (4.2 K); (see also for example [7, 8, 9]).

In the case of metal-organic vapour phase epitaxy (MOVPE) the panorama is more faceted, with a general agreement on the significant complexity of the growth processes [10]. Hence values reported for systems similar to the ones just cited for MBE are somehow lower (e.g. InAs QW in AlSb on GaSb: $\mu_{LT} = 108*10^4$ cm$^2$/Vs) [8]. For this reason, different precursors, carrier gases, growth parameters and group-V switching procedures have been tested to search the highest qualities and performances.

The original motivation of this work was to address the possibility to exploit metamorphic InAs substrates for GaSb/InAs heterostructures, as we had already developed such approach for InAs using $N_2$ as carrier gas. To avoid the intrinsic doping, the relatively high cost and the limited lateral dimensions of real InAs substrates, a specially designed graded layer, progressively changing composition from InP to InAs have been used here as starting pseudosubstrate. Our investigation focused on growths done under nitrogen (as little data could be found in the literature), with an additional interest in the effect of the presence of hydrogen in the carrier gas mixture on the resulting epitaxial layers. We came across a number of relevant issues related to this project and the carrier gas of choice, and we are here extensively reporting on them, despite the fact that we could not fundamentally meet the original ambition. Indeed, we encountered a variety of surprising results related to the usage of nitrogen as carrier gas worth reporting.

Our contribution is structured as follows. For clarity, we deemed necessary to divide the experimental details in two sections. Following the conventional methodology section, a further short description on accurate procedures for XRD investigations and the determination of layer thicknesses is presented. They are relevant for the rest of the discussion. The actual results presentation is divided into five parts. In the first three subsections, we discuss GaSb growth conditions of choice, as preliminary data (i.e., in a way, setting the scene), to introduce the following sections' results in view of (unexpected and unreported) GaSb growth issues, relating to both carrier gas and InAs buffer. In the fourth part we consider the effects of the deposition of InAs on GaSb, revealing a surprising temperature dependent etching/decomposition of GaSb. Finally results of Hall effect measurements for InAs quantum wells are shown, as a quality parameter.

On a side note concerning the carrier gas, it should be mentioned that only a few MOVPE systems worldwide actually run on alternative carrier gases like nitrogen [11, 12, 13, 14] while most systems rely on the well-established $H_2$ [15]. It should also be said, as it is also relevant to our work, that, despite possible merits of alternative carrier gases, there are reports suggesting that $H_2$ favours the decomposition of metalorganics, at least when an atmospheric pressure process is discussed. [16]

## 2. Methodology

**2.1 Growth and characterization.**

All samples were grown in an AIX 200 horizontal 3x2 planetary reactor at 80 mbar using double purified nitrogen [14, 17] with 0-30% hydrogen in nitrogen as carrier gas. All temperatures given are real (estimated) sample temperatures, extrapolated from thermocouple ($T_c$) readings mapped to a calibration curve. Differences (between the $T_c$ reading and the actual sample temperatures) range from ~40 °C at $T_c$ = 590 °C to ~0 °C at $T_c$ = 470 °C. The precursors used for all materials are arsine ($AsH_3$), trimethylantimony (TMSb), trimethylgallium (TMGa), trimethylindium (TMIn) and phosphine ($PH_3$) for the graded substrate buffer.

As substrate we used InAs metamorphic substrates (pseudosubstrates from now on) for all the samples and occasionally an additional manufacturer-bought (001) InAs substrate, in order to compare the influence of both wafer types. The pseudosubstrates were grown in separate runs on semi-insulating (001) InP with ~0.4° misorientation towards [111]B. The structure consists of 100 nm InP buffer, 50 nm InGaAs lattice matched to InP, 1650 nm $In_xGa_{1-x}As$ grading (x = 0.53 to 1.0) and a 100 nm InAs cap. The grading itself consists of three layers changing indium content parabolically to 70% in 400 nm, linearly to 95% in 1000 nm and linearly to 100% in 250 nm, while the temperature is also changed. The final layer of the pseudosubstrates typically shows a small residual strain in the one per thousand range, as expected when metamorphic layers are grown. Further information can be found in earlier publications. [18] Due to the misorientation of the original substrate and the relaxation process the InAs layer grows tilted with an angle of ~1° towards the sample normal and in one of the <110> directions, while showing steps of 2-20 nm height and 0.5-1 µm length resulting in a root mean square (RMS) roughness of approximately 4-6 nm. An atomic force microscopy image can be found in Figure 4a. Further information on the pseudosubstrate will be given in later sections when relevant.

Prior to the growth of the actual investigated samples the substrates were heated to 650 °C under $AsH_3$ for 6 min to perform a deoxidation, followed by the deposition of a 200 nm InAs buffer layer (unless stated otherwise in the text) at 540 °C in 100% $N_2$-carrier gas, which we label here as standard conditions for InAs in our reactor. The best parameters we found for the growth of GaSb (similarly to other reports) were a temperature of T = 550 °C and a V/III ratio of 1, as will be seen from the following discussion. As previously stated, nitrogen is normally used as carrier gas in our system, as other groups do [19, 20, 21]. For reasons which will become clear in the following of the text, we added some amount of hydrogen to the carrier gas mixture, whereas only nitrogen was flowing through the metal-organic sources. We anticipate that with a molar flux of 40 μmol/min for TMGa and TMSb we achieved at best a growth rate of GaSb roughly 100 nm/h. This is much lower than the growth rate we would predict from other materials grown in our reactor, and it is much lower than other comparable reports in the literature (see for example [1, 14, 22, 23]), except for what reported by Agert et al. [24] when nitrogen was used as carrier gas. We will discuss this point further in our text.

The surface morphology was investigated by optical microscopy in Nomarski contrast mode and by atomic force microscopy (AFM) in tapping mode. RMS roughness values were calculated from 10x10 μm$^2$ images after applying a 2$^{nd}$ order flattening. A FEI QUANTA FEG 650 scanning electron microscope (SEM), equipped with an energy dispersive X-Ray spectroscopy system (EDX), was used to investigate the layer thickness. Hall effect measurements were done at room temperature (RT) and in liquid nitrogen (LN) at 77 K at a maximum magnetic field of 0.3T in the Van der Pauw geometry. High resolution X-ray diffraction measurements (XRD) were done with an X-Pert MRD diffractometer using a hybrid monochromator and a triple axis analyzer. Due to the use of the metamorphic buffer layer the investigation of these samples by XRD demands some specific precautions, which are described in the next subsection.

### 2.2. On the XRD analysis

As mentioned above, the samples grown on pseudosubstrates demand some special attention. Due to the miscut of the InP wafer the graded layer is growing tilted on the substrate. As a result of this tilt the lattice planes of the InP substrate and the final InAs buffer layer include an angle of ~ 1° in one of the two <110> directions, and a much smaller one in the complementary <1-10> direction (more details can be found in [18]). In the correct alignment of the sample, the 1° tilt angle appears as an offset in $\chi$ (sample tilt perpendicular to the diffraction plane). A reciprocal space map (RSM) of a representative sample is shown in Figure 1a as contour plot. On the right the strong reflection of the substrate is visible, followed by a broad region on the left, which can be attributed to the metamorphic buffer layer. Further on the left the reflections of the InAs buffer layer and GaSb layer are visible. They have a significant full width at half maximum (FWHM) in $\omega$-direction, which is most probably caused and dominated by the underlying graded buffer layer. The RSM was recorded in alignment to the InAs peak, which means that the InP peak is not perfectly aligned due to the offset in $\chi$ between the InAs buffer layer and the substrate.

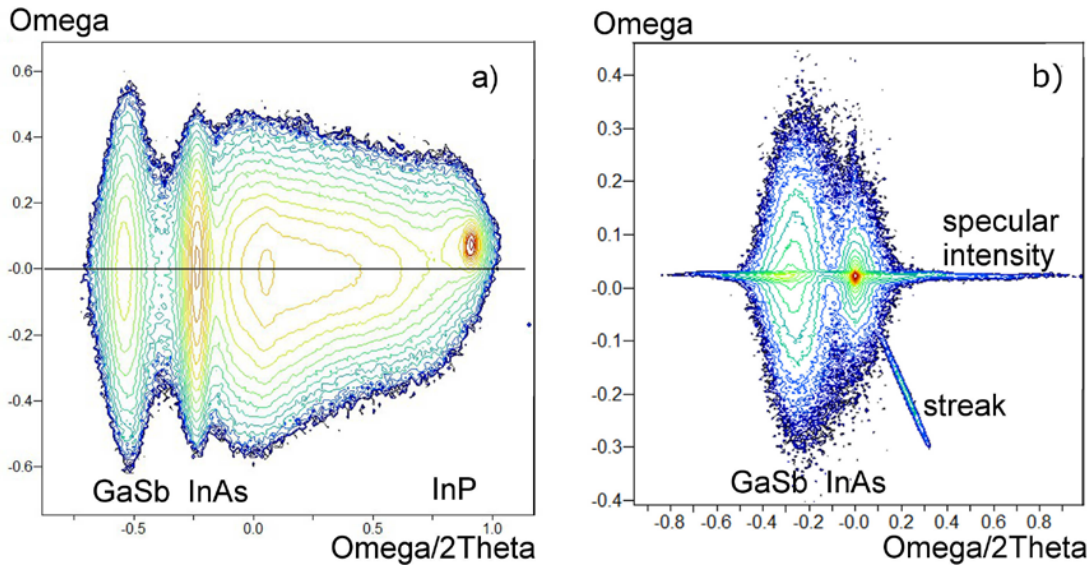

**Figure 1** Reciprocal space maps (contour plot) of 004 reflexion of GaSb on InAs pseudosubstrate (a) and on manufacturer-bought InAs substrate (b). For sample (a) the offset in χ causes a shift between substrate and layer peak regarding ω. The line visualizes that substrate and layer peak are not perfectly visible in a single ω/2θ scan. b) On InAs a monochromator streak and a specular reflected peak are visible over a wide angular region probably due to some mosaicity.

In general, a simple ω/2θ diffraction scan of a misaligned crystal will reveal a false (bigger) lattice constant and an offset in ω as well. The InP-peak appears not at the natural ω, but shifted to slightly higher values. Indeed, recording a single ω/2θ (or 2θ/ω) double rocking curve (DRC) in this alignment, the scan will not necessary hit the maximum of the substrate or the layer peak. For this reason two DRCs were always recorded in alignment, especially regarding χ, one to the InP substrate and the other to the InAs buffer layer and combined in one graph (e.g. see Figure 2a upper curves). When measuring aligned to InAs or GaSb, the InP peak will appear less intense, while in alignment to InP the layer peaks look slightly less intense. Furthermore, when measuring aligned to InP, the position of the peak of the misaligned layer is shifted to lower angles and hence the lattice constant of the misaligned layers seems higher. For this reason, for highest accuracy in the determination of the strain state of the epitaxial layers (via simulation software) we always took care to use the measurement values recorded by the instrument properly aligned to the layer peak.

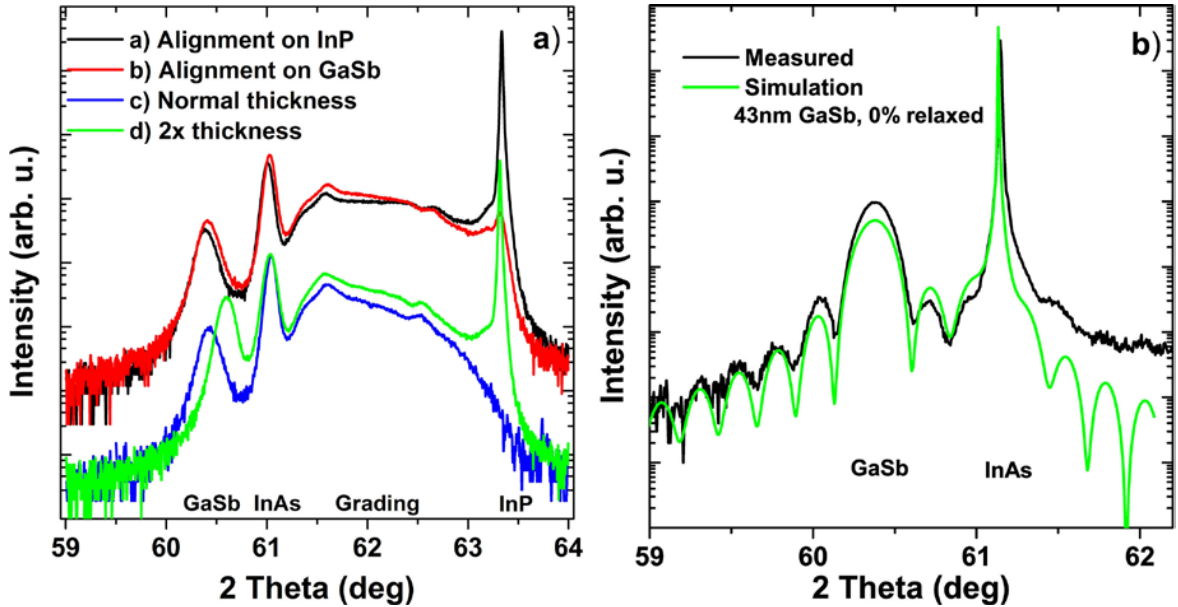

**Figure 2 2θ/ω double rocking curves. a) Samples grown on pseudosubstrate. Upper curves: different alignment of the sample, lower curve: samples with different GaSb layer thickness. b) Sample grown on manufacturer-bought InAs with simulation of diffraction pattern.**

In Figure 1b a RSM of a GaSb sample grown on a manufacturer-bought InAs substrate is shown. In general, compared to the samples grown on the pseudosubstrates, the FWHM in ω-direction is lower, which points to a higher crystalline quality. Nevertheless, there is another feature of the samples grown on manufacturer-bought InAs: namely there is a specular diffraction signal which can be followed over a long angular region in the 2θ/ω-direction. Similar features have been observed for metal films in [25] and for group III-Nitrides [26, 27]. This feature can be explained by a modified mosaic block model [25]. Without discussing deeply the reported model, one of its consequences is the fact that the crystal quality, estimated from the FWHM in ω-direction, is being overestimated. Hence, special attention has to be paid when ω rocking curves of these types of samples are examined as means to judge sample quality.

Another challenge we had to face was the determination of the thickness of the GaSb layers. As already mentioned, the growth rate was (unexpectedly) very low (~ 100 nm/h), so only relatively thin layers could be deposited in a reasonable time with acceptable growth cost. Unfortunately we do not have direct access to transmission electron microscopy (TEM). Attempts to measure the thickness by cross section AFM and cross section SEM were difficult, because of the low contrast between the graded buffer and the InAs and GaSb layers. Furthermore, thickness fringes in XRD DRCs, that can be used for quantitative thickness determination, were only visible for some samples grown on manufacturer-bought InAs substrates (see Figure 2b), while they were missing totally for samples grown on the pseudosubstrates. This might be due to lower quality of GaSb on the grading but it is most probably an induced feature of the stepped surface and the tilted crystal.

Additionally, cross section AFM/SEM suggested that the layer thickness is lower for samples grown on manufacturer-bought InAs. Most probably this is due to a favoured nucleation on the stepped surface of the pseudosubstrates. Hence we caution the reader again, that in general the thickness determination of GaSb layers was not easily established. Thicknesses measured with different methods (e.g. SEM /AFM cross sections and XRD) could give significant error bars and growth rates did not appear necessarily the same on the manufacturer-bought InAs substrate and on the grading. For this reason the given thicknesses of GaSb are only estimates and growth rates could be only determined imprecisely (± 20 %). Hence, only few numbers are given in the following discussion and are based on a qualitative rating. For example, we assumed that for thin layers the integrated intensity of the signal in XRD is strongly dependent on the layer thickness. Furthermore, we expect that the FWHM in $2\theta/\omega$-direction will be smaller for thicker layers as, according to mosaic block models, the vertical coherence length is increased. [28]

A representative example corroborating our procedure is shown by the lower curves in Figure 2a. Both samples were grown with the same growth parameters and differ only in terms of growth time (hence layer thickness). Comparing the GaSb peaks it is obvious that the one of the thicker sample is higher and narrower. The shift in position, which is also visible, will be discussed later on. We also stress that this lack of precision on layer thickness determination does not affect the conclusion of our manuscript, as will be clear in the following of the text.

## 3. Results

As stated before, for all samples discussed in this paper InAs pseudosubstrates were used, while for some experiments additional samples on manufacturer-bought InAs substrates were grown in the same runs. Results from the latter samples will be marked explicitly. Developing the procedure for the GaSb deposition we investigated the influence of temperature, V/III ratio and especially the in influence of the addition of some hydrogen. We observed an unexpected decomposition of GaSb layers after overgrowing them with InAs, and this issue is discussed in the fourth subsection. Finally the results of electrical characterization by Hall measurements are shown.

### 3.1. Preliminary investigations

In the early stages of this research, to develop a reliable process for GaSb growth on InAs by MOVPE with nitrogen as carrier gas, we investigated a broad range of growth conditions for GaSb deposition, using the existing literature as a guideline. We encountered a number of difficulties in developing reliable procedures with pure nitrogen as carrier gas, and in the process we investigated the addition of hydrogen in the carrier gas mixture, with some success. This development has not been particularly linear in its structure, and we summarise it here for clarity only as introductory step to the rest of our work: starting at the highest temperature of 630 °C, V/III ratios 1-2 and a carrier gas mixture of maximum 10-20% $H_2$ in nitrogen, we only observed a broad peak in XRD 2θ/ω curves at high angles and a roughening of the surface (both not shown). The feature in XRD could be attributed to GaAsSb, but not to the aimed-at pure GaSb which would have appeared at much lower angles. The GaAsSb signal indicated some interface reactivity and decomposition of InAs, since no $AsH_3$ was supplied during the growth. Lowering growth temperatures, we could observe a signal at a position compatible with a partially strained GaSb from 570 °C on, independent of the V/III ratio used when < 4. Best results (in this preliminary investigation) were achieved at 550 °C and a V/III ratio of 1, which indeed resemble values reported in the literature (see for example [1, 4, 7, 22, 23, 29, 30, 31, 32, 33, 34, 35, 36]). For temperatures lower than 550 °C the surface roughness became higher, while no clear trend was visible from XRD measurements. Lowering the V/III ratio to 0.6-1 had no influence on samples grown on the graded buffer, but led to a broadening of the XRD peak of GaSb on manufacturer-bought InAs substrates in the ω-direction, which points to a potentially reduced crystal quality (not shown).

### 3.2. The influence of hydrogen

These preliminary results, together with some discussions with other colleagues [37], reinforced the idea that the carrier gas mixture could have a (positive) role in determining the quality of GaSb growth. It should be mentioned that also in other material systems, such as InN, a similar behaviour has been observed [38]. Indeed, in III-V MOVPE growths performed using a hydride group V precursor, extra atomic hydrogen is provided by group V precursors such as $AsH_3$ or PH$_3$ [16], but is missing when only TMSb is used. Moreover reports suggested that hydrogen is essential for the effective decomposition of the metal organics. [39]

For these reasons, in the following (for the entire length of the growth process) nitrogen as carrier gas was investigated with a fraction of hydrogen, with the intent to support the decomposition of the metalorganics, when growing GaSb and InAs heterostructures. In Figure 3 the influence of increasing the hydrogen content in the carrier gas on the XRD pattern of a GaSb buffer layer is displayed. All samples were grown in identical conditions concerning temperature, V/III and layer growth time (550 °C, 1, 36 minutes) on our standard pseudosubstrates. The FWHM of the GaSb peak is continuously decreasing with increasing H$_2$ content at constant carrier gas and total molecular flow (diamonds), while the total peak height is increasing (not shown). Both are signs of an increase in layer thickness, according to the mosaic block model [40] and the discussion presented in section 2.2. Also a shift of the peak position can be observed, which we attributed to a change in the strain state (squares), coherently with the idea that for thin, strained layers the degree of relaxation will increase with the layer thickness.

This effect was tested and verified by the growth of a sample with 10% H$_2$ and doubled growth time, i.e. thickness (open square), which indeed showed a higher relaxation. DRCs of these two samples are displayed as lower curves in Figure 2a. In our data interpretation, and since the amount of precursors supplied was the same for all samples, this leads to the conclusion that the growth of GaSb is possible with nitrogen only, but is more efficient in the presence of hydrogen.

It should also be said that, in general, the roughness of samples grown on the pseudosubstrate (RMS = 15-20 nm) was not influenced by the H$_2$ content, and fundamentally follows the roughness of the underlying pseudosubstrate. The surface morphology of samples grown on manufacturer-bought InAs appeared to have a slightly lower roughness when grown with a $H_2$ content of 10%. Furthermore, the 10% $H_2$ samples on manufacturer-bought InAs showed most pronounced thickness fringes, indicating a layer thickness of 40 nm (Figure 2b) for this particular sample. Therefore, to limit the amount of variables, we focused our further investigations on samples grown with carrier gas mixture containing 10% hydrogen.

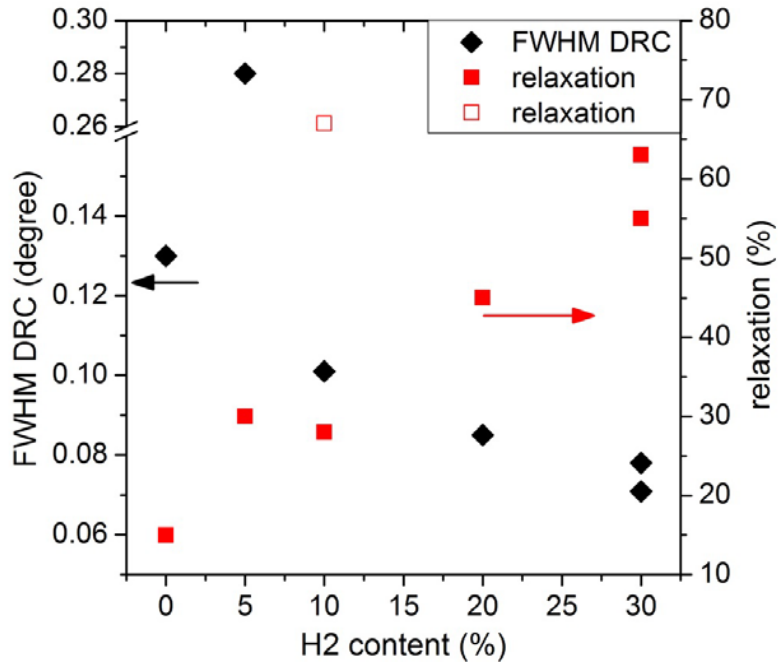

**Figure 3 Influence of the $H_2$ content in carrier gas on the GaSb XRD peak. FWHM in 2θ/ω-direction and the degree of relaxation calculated from the peak position. Sample with doubled thickness is marked by the open symbol.**

### 3.3. Substrate/InAs buffer layer thickness role

One of the surprising issues has been the effect of the InAs buffer layer thickness, which was varied between 50 and 500 nm, depending on the sample. The biggest effect was encountered when growing GaSb on manufacturer-bought InAs wafers: we found rectangular, elongated hillocks aligned along one of the InAs [110] directions (see Figure 4c, d). Remarkably, we never observed these elongated hillocks on the pseudosubstrates, but exclusively when we used the InAs wafers (an unexpected advantage of the pseudosubstrate).

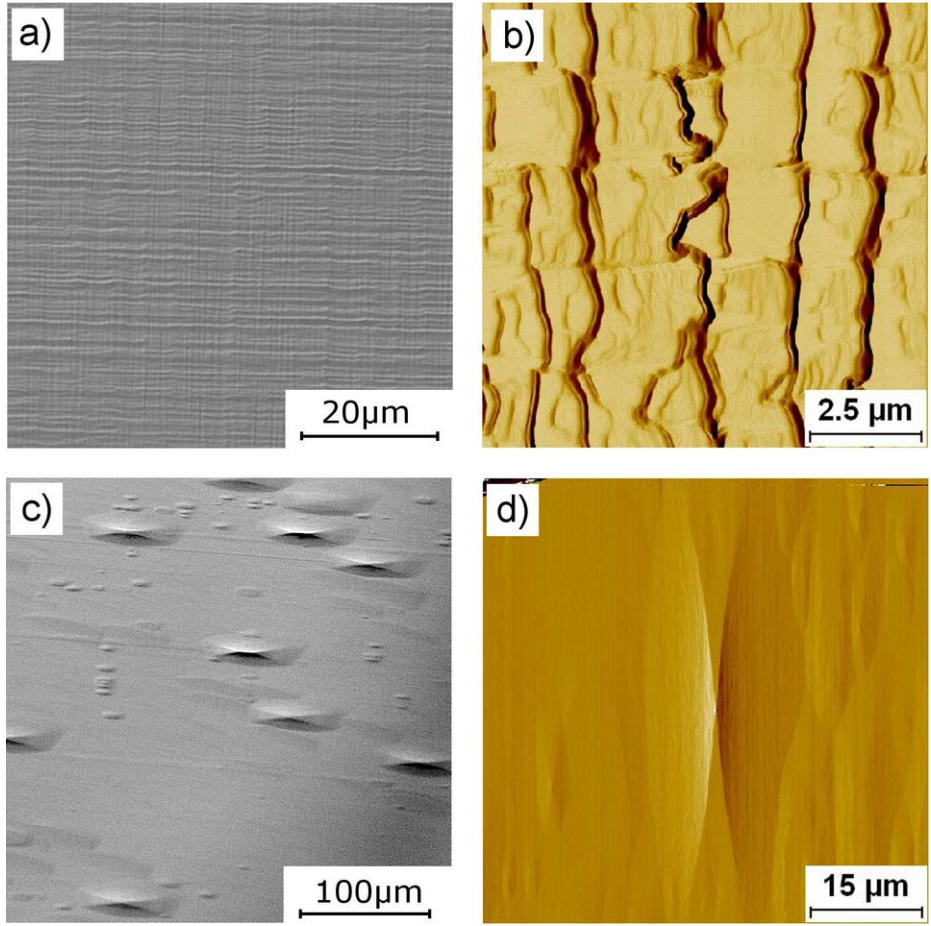

**Figure 4** Microscopy and AFM amplitude signal images of a) the InAs pseudosubstrate (RMS ~ 5 nm), b) GaSb layer grown on pseudosubstrate (RMS ~ 15 nm), showing no elongated hillock formation, c, d) GaSb layer grown on manufacturer-bought InAs substrate (RMS <1 nm) showing the appearance of elongated hillocks.

These hillocks had dimensions of up to 70x20 $\mu m^2$, a significant height of up to 250 nm and consist of GaSb, as confirmed by EDX measurements. We stress that the hillocks appear significantly pronounced in height, when compared to the relatively small amount of deposited material. Nevertheless, increasing the thickness of the InAs buffer layer from 50 to 200 and 500 nm the density of these hillocks dropped from a nearly full coverage to $< 30/mm^2$. The influence on the X-Ray ω-rocking curves of the GaSb 004 reflection is shown in Figure 5. Depending on the orientation of the hillocks to the diffraction plane, the ω-rocking curve looks different for samples with thin (50 nm) InAs buffer layer (Figure 5a). In perpendicular alignment the curve is broad and has a Gaussian shape, while in parallel alignment the peak is narrower and consists of a convolution of a Gaussian and a sharp Lorentzian peak. With increasing InAs buffer thickness (Figure 5 panels b, c) both orientations become similar, which somehow points to an increased homogeneity. Also the overall FWHM is lower, which hints to a higher crystal quality (keeping in mind limitations given in paragraph 2.2).

It is also noteworthy that for samples grown on the grading we could not observe an effect on the FWHM. However, the roughness of the GaSb layer was reduced (from 20 to 10 nm, not shown) by increasing the buffer layer thickness from 50 to 500 nm, and indication perhaps that the actual buffer thickness still bears a (surprising in its extent) role.

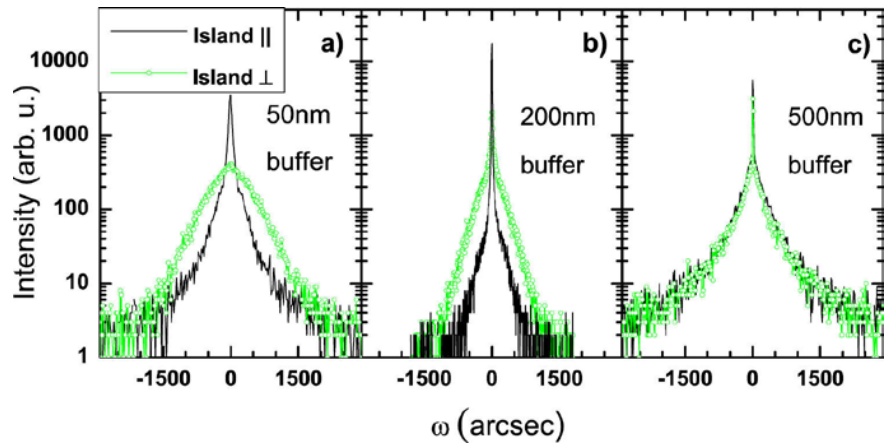

**Figure 5 ω-rocking curve of the GaSb (grown on commercial InAs substrates) peak for an alignment of the islands parallel and perpendicular to the diffraction plane. With increasing InAs buffer thickness (a-c) the Gaussian background is reduced and both orientations become indistinguishable.**

### 3.4. InAs (QW) in GaSb barriers

The results reported in sections 3.1-3.3 gave us important grounds for the following of our investigations. After developing a reliable process for GaSb with nitrogen as main carrier gas and "understood" associated issues, we subsequently studied InAs/GaSb heterostructures, and only InAs pseudosubstrates were used. On the pseudosubstrates the samples consisted of a 300 nm InAs buffer layer, a GaSb layer of an estimated thickness of 70 nm, and a nominally 20 nm InAs quantum well (QW), which was covered by ~25 nm GaSb in the final stage. For all samples $AsH_3$ is supplied during a 180 seconds temperature ramp between the InAs buffer layer and the GaSb layer. Nevertheless, before discussing the results concerning the full structure in section 0, we need to address in this section more surprising peculiarities associated with the growth of InAs on GaSb.

Selected AFM images are shown in Figure 6 and XRD double rocking curves (DRC) are shown in Figure 7, with curve *a)* as reference (GaSb on InAs buffer only). We apologize for the apparent lack of optimal order of samples in Figure 7, which was chosen for a clearer overall appearance.

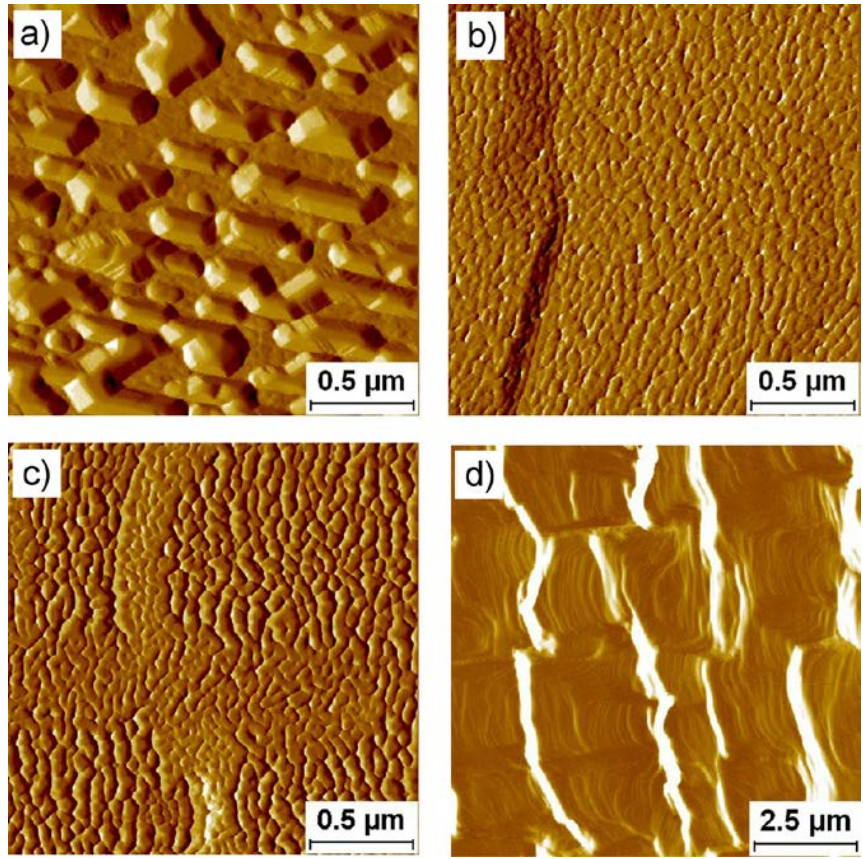

**Figure 6** Representative AFM amplitude signal images of a) 20 nm InAs on GaSb grown at 550 °C, b) GaSb after 60 s AsH$_3$ supply, c) GaSb after 60 s TMSb supply. d) full structure of InAs QW (520 °C) and GaSb cap (550 °C), (RMS 5-10 nm). In panel d) a heavy step bunching of crystallographic steps is evident (whiter regions).

**Decomposition issues:** After growing 20 nm InAs at 550 °C on top of GaSb (see also Figure 6a) we observed a surprising reduction of the GaSb peak height and area in XRD measurements displayed in Figure 7 Curve *d)*. Additionally a shift towards a more strained status was observed, even though the underlying structure was the same as the reference (comparing Curves *a)* and *d)*). This feature can only be explained by a decomposition or etching of GaSb during the InAs growth (duration 70 s), as we did not observe alloyed peaks appearing. For this reason, further experiments were needed to investigate the factors promoting the decomposition, for example by observing the effects of supplying different precursors after the GaSb layer growth at a temperature of 550 °C.

Indeed after 60 seconds of AsH$_3$ supply the surface looked somewhat rippled, as shown in Figure 6b, whereas the XRD peak (Curve *e)*) was just slightly less intense than the reference. A similar reordering was also found after 60 seconds of simple TMSb supply (Figure 6c). The XRD peak was lower, broader and shifted towards a more relaxed position (Curve *f)*). During this step some kind of reordering must have happened in which strongly relaxed GaSb formed the rippled structure. This could be explained by an As for Sb exchange reaction, as also reported for example in [41]. The strongest effect nevertheless was seen by the supply of TMIn, since the GaSb peak vanished totally after 30 seconds (Figure 7, Curve *g)*). Most probably gallium was replaced by indium, forming InSb which must have decomposed (and disappeared) at elevated temperatures [42,43], as no sign of extra peaks and features in the XRD spectrum is evident. This process seems to be also happening during the InAs deposition and could explain the observed decrease of the GaSb signal. We also found that these decomposition processes are strongly temperature dependent. For example the morphology and the appearance in XRD of a bare GaSb layer is unaffected if it is exposed to AsH$_3$ at 520 °C for ~ 120 seconds and if AsH$_3$ or TMSb is supplied during immediate cool down from 550 to 350 °C (temperature, at which we stopped supplying the group V precursor). Also samples which were overgrown with InAs at lower temperatures (we tested 470, 500, 520 °C), do not show signs of decomposition, referring to an unchanged GaSb signal in XRD compared to the reference sample.

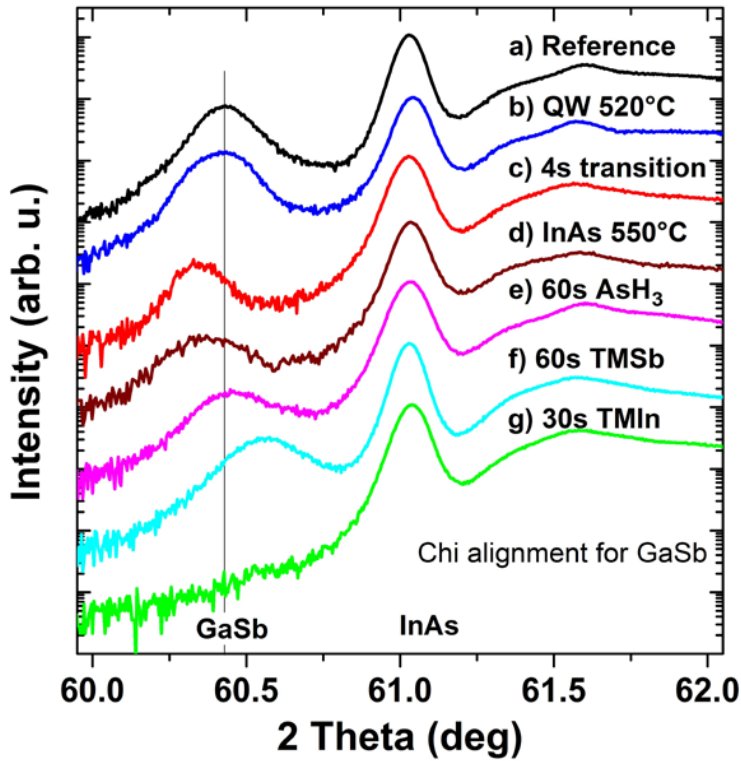

**Figure 7** 2θ/ω double rocking curves of GaSb on InAs pseudosubstrate (intensity scaled and order of samples chosen for clear view). a) Reference: 70 nm GaSb on InAs pseudosubstrate, b) InAs QW grown at 520 °C, capped with GaSb, c) uncapped InAs QW grown at 550 °C after 4 s transition (AsH$_3$ + TMSb), d) InAs directly grown on GaSb at 550 °C, e-g) decomposition study at 550 °C, GaSb layer exposed to: e) 60 seconds AsH$_3$, f) 60 s TMSb, g) 30 s TMIn.

**QW growth:** When we started with a direct overgrowth of InAs on GaSb at a temperature of 550 °C, we observed the formation of elongated hillocks, which volume roughly matches that of the layer's aimed thickness (see Figure 6a and discussion in section 3.3). At this point we want to stress that 550 °C is similar/close to our "normal" InAs growth temperature (540 °C) which gives no issues, when grown directly on InAs. To influence and favour the nucleation of InAs on GaSb we tried also to supply both TMSb and AsH$_3$ for only 4 seconds prior the InAs deposition at 550 °C, which we term a transition step. By this step the islands became more closely spaced (not shown) and the GaSb signal maintained stronger in the DRC shown in Figure 7, Curve *c)*.

Comparing this (Curve *c)*) with the InAs layer without transition (Curve *d)*), one can see that the GaSb peak is stronger but still shifted and lower in intensity than the reference (Curve *a)*). It indicates that GaSb must still have decomposed a little while the 2$^{nd}$ InAs layer was deposited but the transition step prevented a strong decomposition of GaSb. This trend of protection of GaSb and improved InAs morphology was confirmed for a sample with a 12 seconds transition step (not shown). Furthermore, we found that InAs QW layers grown at lower temperatures ($\leq$ 520 °C) do not show islanding nor strong morphology changes, but atomic steps on the terraces, following the underneath grading. Also, the morphology is independent from whether InAs is grown directly or a transition step is added, supplying $AsH_3$ + TMSb or $AsH_3$ + TMSb + TMIn as somehow discussed in [35] too.

To at last create an optimised 2D electron gas QW-structure, not intentionally doped InAs films were overgrown with a ~ 25 nm GaSb layer. As shown in Figure 4b and 6d the morphology of the full QW structure resembles the underlying GaSb layer and the X-Ray signal is similar to the reference (Figure 7, Curve *a)* and *b)*). For the samples used in the next section 3.5 the following standard parameters have been used: temperature ramp after GaSb layer under TMSb, 4 seconds transition, InAs QW at 520 °C, temperature ramp after InAs QW under $AsH_3$ and GaSb capping layer at 550 °C. Furthermore, for completeness, we investigated samples which were grown at different temperatures for the QW (450-550 °C) and the GaSb cap layer (520-550 °C). In general the morphology was unaffected while the influence on the results of Hall measurements is shown in Figure 8 for selected samples.

### 3.5. Hall structures

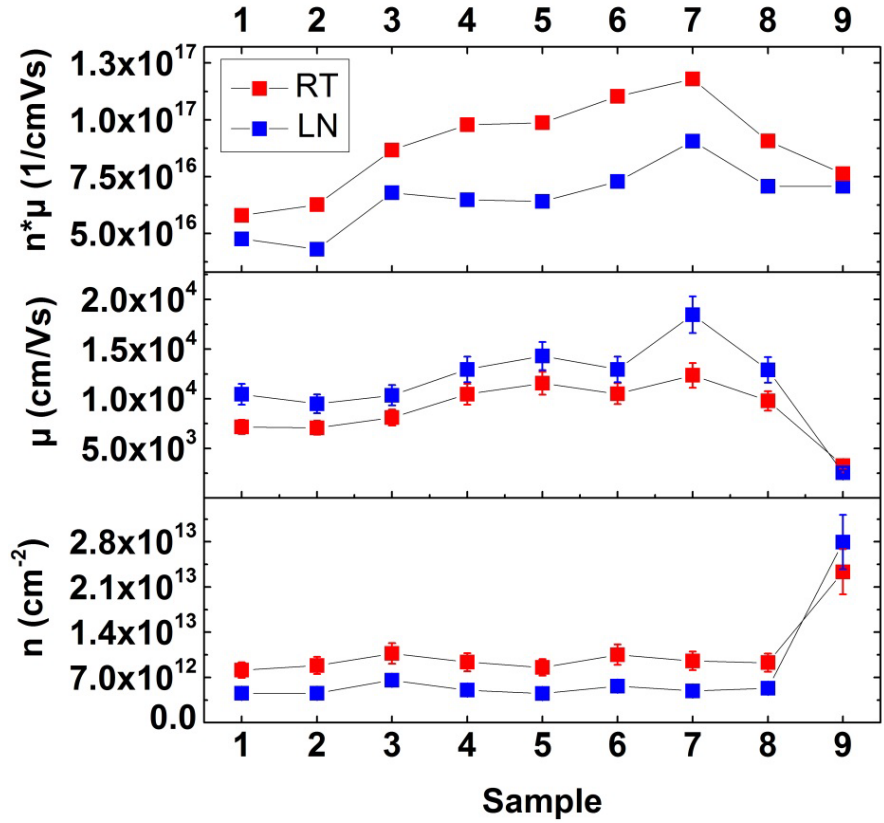

**Figure 8** Values of electrical measurements of various samples with increasing number of layers. Initial material studies: 1) pseudosubstrate only: grading + 300 nm InAs, 2) pseudosubstrate + 80 nm GaSb, 3) pseudosubstrate + 80 nm GaSb + 20 nm InAs; Full QW structures containing pseudosubstrate, 80 nm GaSb buffer, InAs QW of given thickness and 25 nm GaSb cap: 4) 20 nm InAs QW, 5) 40 nm InAs QW (InAs QW grown at 520 °C, temperature ramp after InAs QW under $AsH_3$ and GaSb capping layer at 550 °C); Full QW structures containing pseudosubstrate, 80 nm GaSb buffer, 20 nm InAs QW and 25 nm GaSb cap grown in various growth conditions: 6) InAs QW grown at 490 °C, GaSb capping at 550 °C; 7) InAs QW and GaSb cap grown at 520 °C, 8) GaSb cap grown after additional 6min growth interruption under $AsH_3$, 9) reference: sample grown with $N_2$ only in standard conditions.

In Figure 8 results of Hall measurements (Van Der Pauw geometry) of various samples are discussed: the sheet carrier density (n), the mobility (μ) and the product of n*μ measured at room temperature (RT) and at liquid nitrogen temperature (LN) are displayed. The results for different samples are marked by numbers on the x-axis. Error bars are estimated to be ±15% for n and ±10% for μ due to scattering of measurements for different sample regions and contact quality. As reported by authors in [44, 45, 46] for example, InAs shows a surface accumulation of carriers which forms a two dimensional electron gas (2DEG) at all heterogeneous interfaces, which we exploit for obtaining a conductivity in our GaSb/InAs/GaSb electron confining structures which are believed to be type II like [47, 48]. The surface accumulation contributes to the overall conductivity, in form of two dimensional electron gases at all InAs interfaces, which to some extent complicates an exact, exclusive determination of the properties of an InAs QW. Normally, for all our samples, the carrier densities are in the $5*10^{12}$ cm$^{-2}$ or $9*10^{12}$ cm$^{-2}$ region for LN or RT measurements, respectively. The mobility, on the other hand, shows some dependence on the detailed structure. Similar mobilities, which are most probably dominated by the surface accumulation layers, are observed for the simple InAs pseudosubstrate only (1), the sample with GaSb cladding layer (~ 70-80 nm) on the pseudosubstrate (2) and the sample with an uncovered InAs QW (3) on top of the latter sample. When the QW is overgrown by a ~ 25 nm GaSb layer (4) the mobility is increased, whereas the thickness of the QW (5) has some influence as well. An explanation for these features could be a difference in confinement and the role of the stepped surface, which would cause thickness fluctuations as well, somehow affecting the electron mobility. Furthermore, one can notice that the difference between RT an LN values for the product n*µ is bigger for the last two samples discussed. Samples with an InAs QW grown at lower temperatures show a slightly lower mobility (6) so that the best growth temperature we investigated is 520 °C. An improvement regarding the mobility can be seen for Sample (7), where the GaSb cap was grown at 520 °C immediately after the QW growth. This improvement can be due to a better quality of the GaSb cap or some unfavourable effect on the QW during the heat up to 550 °C under arsine from the QW growth temperature.

To check for the influence of AsH$_3$ an InAs QW was exposed to AsH$_3$ for 6 min at 520 °C before it was capped by GaSb (8). The mobility was lower again, which points to structural changes of the InAs QW during the time the temperature is increased, for example for Sample (6). Sample (9) was grown with nitrogen as carrier gas only as a further test of the importance of hydrogen and shows a quite different behaviour. The mobility is much lower, while the carrier density is 3 times higher. This points to a layer with more defects providing the high number of carriers. Also the morphology is worse and less uniform, which shows that at least some hydrogen is essential for the growth of good quality GaSb. It should be nevertheless noticed that for sample (9) no attempt to correct for actual growth rates was done, and our observations should be considered only as indicative.

## 4. Conclusion

In conclusion, we have shown that it is possible to substitute InAs wafers with InAs metamorphic substrates during the growth of the InAs/GaSb system by MOVPE, using a nitrogen/hydrogen mixture as main carrier gas, obtaining two dimensional electron gas structures with unintentional (but expected) n-charging up to $9/5*10^{12}$ cm$^{-2}$ and mobilities up to $1.2/1.8*10^{4}$ cm/Vs at room and liquid nitrogen temperature respectively. We nevertheless discussed a number of peculiarities and unexpected results which need to be considered to achieve good quality epitaxial growth under such conditions. GaSb growth by MOVPE on InAs wafers and InAs metamorphic substrates is possible in a relatively narrow range of growth parameters. When nitrogen is used as carrier gas, growth occurs in a regime of low effective growth rates and benefits strongly from a mixed $N_2/H_2$ environment. The effective InAs buffer layer thickness also strongly influences growth quality, while the growth of InAs on GaSb, with nitrogen as carrier gas, can result in unexpected GaSb etching, which is active even at moderate growth temperatures, but disappears at sufficiently low ones. We believe our findings will be useful to the crystal growers' community, giving first indication on how this complex material system can be managed by MOVPE when nitrogen is used as main carrier gas.

**ACKNOWLEDGMENT**

This research was enabled by the Irish Higher Education Authority Program for Research in Third Level Institutions (2007-2011) via the INSPIRE Programme and partly by Science Foundation Ireland under the IPIC award 12/RC/2276 and grant 10/IN.1/I3000. We also gratefully acknowledge Intel Corporation for financial support.

We thank Prof. Hilde Hardtdegen for useful discussions and suggestions and Dr. K. Thomas for the MOVPE system support.

[15] We realised in these years that, for some reason, the MOVPE community appears strongly polarised on the subject of carrier gas choice. We are not. We mention here only for the sake of completeness that $H_2$ has been historically the only properly economical "purifiable" option for many years, and that our laboratory chose $N_2$ simply because it involved smaller equipment cost at purchasing time (e.g. no hydrogen line, no hydrogen purifier, no hydrogen sensors): it was the only way to match our equipment budget. It should also be mentioned that the running costs for an MOVPE kit are, in general and all considered, comparable for $N_2$ and $H_2$ operated systems in terms of consumables (if run at low pressure), as slightly lower growth rates when $N_2$ is utilised (therefore higher hydrides consumption, a significant fraction of consumable costs) can often be

roughly compensated and offset by the price difference between nitrogen and hydrogen, again all considered. The overall financial balance will depend on the usage of the installed kit, source of nitrogen (bottled or from liquid nitrogen tank), sharing of the hydrogen generator with other on site facilities, etc. Obviously, hydrogen being an explosive gas, is also higher in safety concerns in comparison with neutral nitrogen: this contributes to the laboratory budget. The interested reader might find of use to refer to, for example, to [11] and/or [21].